# Partial Hadamard Encoded Synthetic Transmit Aperture for High Frame Rate Imaging with Minimal $l_2$-Norm Least Square Method

Jingke Zhang[1, #], Jing Liu[2, #], Wei Fan[2], Weibao Qiu[3, 4, *], and Jianwen Luo[1, *]

[1] Department of Biomedical Engineering, School of Medicine, Tsinghua University, Beijing 100084, China

[2] Shenzhen Mindray Bio-Medical Electronics CO., LTD, Shenzhen 518057, China

[3] Paul C. Lauterbur Research Center for Biomedical Imaging, Shenzhen Institutes of Advanced Technology, Chinese Academy of Sciences, Shenzhen 518055, China,

[4] Shenzhen Key Laboratory of Ultrasound Imaging and Therapy, Shenzhen 518055, China.

[#]These authors contributed equally to this work

Corresponding authors: luo_jianwen@tsinghua.edu.cn and wb.qiu@siat.ac.cn

**Abstract:**

Synthetic transmit aperture (STA) ultrasound imaging is well known for ideal focusing in the full field of view. However, it suffers from low signal-to-noise ratio (SNR) and low frame rate, because each array element must be activated individually. In our previous study, we encoded all the array elements with partial Hadamard matrix and reconstructed the complete STA dataset with compressed sensing (CS) algorithm (CS-STA). As all the elements are activated in each transmission and the number of transmissions is smaller than that of STA, this method can achieve higher SNR and higher frame rate. Its main drawback is the time-consuming CS reconstruction. In this study, we accelerate the complete STA dataset reconstruction with minimal $l_2$-norm least square method. Thanks of the orthogonality of partial Hadamard matrix, the minimal $l_2$-norm least square solution can be easily calculated. The proposed method is tested with simulation data and experimental phantom and *in-vivo* data. The results demonstrate that the proposed method achieves ~$5\times10^3$ times faster reconstruction speed than CS algorithm. The simulation results demonstrate that the proposed method is capable of achieving the same accuracy for STA dataset reconstruction as conventional CS-STA method. The simulations, phantom and *in-vivo* experiments show that the proposed method is capable of improving the generalized contrast-to-noise ratio (gCNR) and SNR with maintained spatial resolution and fewer transmissions, compared with STA. In conclusion, the improved image quality and reduced computational time of LS-STA pave the way for its real-time applications in the clinics.

***Keywords*:**

high frame rate, minimal $l_2$-norm least square, synthetic transmit aperture, ultrafast ultrasound imaging.

# I. Introduction:

Synthetic transmit aperture (STA) is a widely adopted technology in the field of pulse-echo imaging, such as radar and sonar systems [1, 2], nondestructive testing [3, 4] and medical ultrasound imaging [5-7], because it is capable of achieving high spatial resolution over the entire imaging area. Complete STA dataset consists of backscattered echoes corresponding to each transmit element and receive element pair. By applying appropriate delay to each backscattered echo and coherently summing the delayed echoes, it is able to synthetically focus at every location, thereby, achieving ideal focusing throughout the field of view. Nevertheless, STA has two main drawbacks. The first one is low signal-to-noise ratio (SNR), which is the consequence of low ultrasound energy carried by single activated element in each transmission. The second one is low frame rate. Because each element must be activated individually in STA, the number of transmissions equals the number of array elements to complete the acquisition of STA dataset. Therefore, the frame rate is limited by the large number of transmissions.

Researchers have proposed many methods to solve these problems. O'Donnell *et al* applied a temporally extended pulse to drive the elements to transmit ultrasound wave [8]. This elongated pulse would increase the incident ultrasound energy to the imaging object and therefore improve the SNR. This method needs to perform decoding process, which would produce significant range side lobe. Researchers need to balance the SNR improvement and range side lobe [8].

Karaman *et al* activated multi-elements to emulate diverging wave from a virtual source behind array [9]. This method is also called multi-element STA. Compared with single-element transmissions, multi-elements transmissions carry higher ultrasound energy. Therefore, multi-element STA is capable of improving the SNR of STA. However, this would reduce the angular spread of diverging transmit beam [10]. Different from multi-element STA which transmits diverging wave, Bae and Jeong [11] activated multiple elements to transmit focused wave and viewed the focal point as a virtual source. Then, the focused wave was considered as two diverging waves spread upward and downward from the virtual source respectively. In the beamforming step, the image area is divided into two parts and each is processed by STA beamforming. Because ultrasound energy of focused wave is very high, this method could improve the SNR significantly. One drawback of this method is that each focused beam only covers a limited spatial region near the focal depth and therefore would introduce a discontinuity near the focal depth. Such discontinuity can be hidden by placing the focal point over the imaging depth [12].

Although the above methods obtain ultrasound image with higher SNR, they cannot acquire the complete STA dataset simultaneously. As a consequence, their focusing is not as ideal as STA. To obtain the complete STA dataset with higher SNR, spatial coding methods such as Hadamard encoding [13-15], S-sequence [16], and delay-encoded transmission [17, 18] were proposed. In these methods, all elements are encoded and activated in each transmission. For an array with N elements, researchers can reconstruct the complete STA dataset after N encoded transmissions by multiplying the backscattered echoes with the inverse of the coding matrix. The theoretical SNR improvement of the Hadamard encoding method is $\sqrt{N}$. However, to guarantee the existence of inverse of the coding matrix, this method needs N transmissions and therefore its frame rate is the same as that of STA. By constructing the linear relationship between the complete STA dataset and dataset from focused wave in the frequency domain, Bottenus recovered the complete STA dataset with improved SNR as well [10]. However, the increase on the frame rate brings by this method is limited.

Hadamard encoding has also been used to obtain other datasets with higher SNR. For example, Misaridis *et al* used Hadamard matrix to encode dataset of sparse transmit aperture [19]. Gong *et al* used Hadamard matrix to encode diverging waves from sub-elements [20]. Tiran *et al* utilized Hadamard matrix to encode plane waves of different steering angles by temporal coding [21]. After decoding by multiplying the inversion of Hadamard matrix, all the above methods can achieve the corresponding dataset with higher SNR.

To obtain the full STA dataset with higher SNR and higher frame rate at the same time, in our previous study, we proposed compressed sensing (CS) based STA (CS-STA) [22, 23]. By encoding the elements with uniformly random matrix and activating all the elements in each transmission, we reconstructed the complete STA dataset with fewer transmissions by utilizing CS algorithm [24]. In our follow-up study, we found that partial Hadamard matrix performs better than uniformly random matrix in terms of image quality and reconstruction time [25, 26]. However, the reconstruction time is still tens of minutes because of the iterative nature of CS optimization algorithms, which prevents it from real-time imaging.

To speed up the reconstruction process, we reconstruct the complete STA dataset with minimal $l_2$-norm least square method in this work. Because of the orthogonality of partial Hadamard matrix, the minimal $l_2$-norm least square solution can be easily calculated. This method was tested with simulation data, experimental phantom and *in-vivo* data. It is shown in Section IV, that the proposed method achieves ~$5\times10^3$ times faster reconstruction speed than CS algorithm, with the same reconstruction accuracy. In addition, the proposed method achieves significant improvement in reconstructed image quality, compared with STA.

This paper is organized as follows: Section II presents the theoretical basis of the proposed method. Section III describes the simulation and experimental setups. Results are presented in Section IV. Section V discusses the results and demonstrates the limitation and future work. Section VI concludes this work.

## II. Theory:

In apodized plane wave (PW) imaging, all the transducer elements are activated simultaneously with different apodizations to acquire Hadamard encoded PW dataset, as shown in Fig. 1(a). In STA imaging, the transducer elements are activated sequentially to acquire STA dataset, as shown in Fig. 1(b). In both PW and STA imaging, full aperture is used to receive the backscattered echoes.

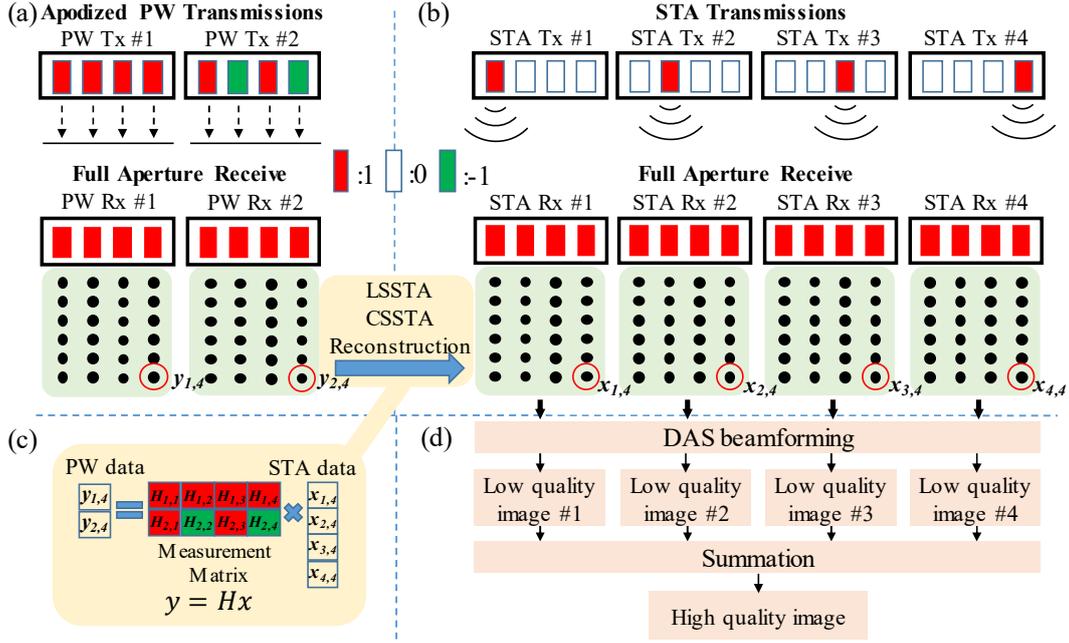

Fig. 1. Diagrams for (a) PW and (b) STA data acquisition. The rectangles denote transducer elements. The different colors correspond to the apodizations applied to the elements. The red circles indicate the slow-time data of PW acquisitions ($y$) and STA acquisitions ($x$), which can be linked with a linear measurement model, as shown in (c). (d) Diagram for delay-and-sum based STA beamforming.

We assume that $x_{n,j}(t)$ denotes the backscattered echo received by the $j^{th}$ ($1 \leq j \leq N$) element when the $n^{th}$ ($1 \leq n \leq N$) element is activated in STA imaging, and $y_{m,j}(t)$ denotes the backscattered echoes received by the $j^{th}$ element for the $m^{th}$ ($1 \leq m \leq M$) transmission in PW imaging. According to linear acoustic theory, the received echoes for a PW transmission is the linear combination of the received echoes of all STA transmissions, and the linear combination coefficients are the apodizations used in the PW transmission [22]. Therefore, there is a linear relationship between $y_{m,j}(t)$ and $x_{n,j}(t)$ as follows [22]:

$$y_{m,j}(t) = \sum_{n=1}^{N} \mathbf{H}_{m,n} x_{n,j}(t) \tag{1}$$

in which $\mathbf{H}_{m,n}$ denotes the encoding amplitude (i.e., transmit apodization) for the $n^{th}$ element in the $m^{th}$ PW transmission. $\mathbf{H}$ is the corresponding encoding matrix. A four-element array is used as an example to demonstrate the relationship, as shown in Fig. 1(a)-(c).

Omitting the time symbol $t$ and denoting $\mathbf{y}_j = [y_{1,j}, y_{2,j}, \cdots y_{M,j}]^T$ and

$\mathbf{x}_j = [x_{1,j}, x_{2,j}, \cdots x_{N,j}]^T$, we obtain the following system of linear equations:

$$\mathbf{y}_j = \mathbf{H}\mathbf{x}_j \tag{2}$$

in which $\mathbf{H} \in R^{M \times N}$ ($M<N$), $\mathbf{y}_j \in R^{M \times 1}$ and $\mathbf{x}_j \in R^{N \times 1}$. $M$ is the number of encoded PW transmissions. $N$ is the number of STA transmissions.

As a result, the problem of obtaining the complete STA dataset from fewer encoded PW transmissions turns out to be solving the linear under-determined problem as denoted by (2) and shown in Fig. 1(c). In our previous study, we assumed that $\mathbf{x}_j$ could be sparsely represented on a sparse basis $\Psi$, i.e., $\mathbf{x}_j = \Psi \mathbf{v}_j$ and most entries of $\mathbf{v}_j$ are close to zero. By solving the following optimization problem with CS algorithm, we can obtain the sparse representation $\mathbf{v}_j$ and therefore obtain $\mathbf{x}_j$ [24].

$$\hat{\mathbf{v}}_j = \underset{v_j \in R^n}{\arg\min} \|v_j\|_1 \quad s.t. \|y_j - \mathbf{H}\Psi v_j\|_2 \leq \varepsilon \tag{3}$$

By choosing the partial Hadamard matrix as the encoding matrix $\mathbf{H}$, we achieved higher image quality and reconstruction speed than uniformly random matrix [25, 26]. The partial Hadamard matrix is the first $M$ ($M<N$) rows of an $N$-order full Hadamard matrix, which is formed as below:

$$\mathbf{H}_{2^1} = \begin{bmatrix} +1 & +1 \\ +1 & -1 \end{bmatrix} \tag{4}$$

$$\mathbf{H}_{2^n} = \mathbf{H}_2 \otimes \mathbf{H}_{2^{n-1}} \tag{5}$$

in which $\otimes$ denotes the Kronecker product. The full Hadamard matrix has a good property of orthogonality, i.e., $\mathbf{H}_{2^n}\mathbf{H}_{2^n}^T = \mathbf{H}_{2^n}^T\mathbf{H}_{2^n} = 2^n I_{2^n}$. That is, both the rows and the columns of the full Hadamard matrix are orthogonal.

For under-determined linear problem such as (2), if the encoding matrix $\mathbf{H}$ is row full rank, i.e., rank($\mathbf{H}$) = $M$, its minimal $l_2$-norm least square solution is obtained by solving the following problem [27]:

$$\min \|\mathbf{x}_j\|_2^2 \quad s.t. \, \mathbf{y}_j = \mathbf{H}\mathbf{x}_j \tag{6}$$

This problem has the following analytical solution:

$$\hat{\mathbf{x}}_j = \mathbf{H}^T(\mathbf{H}\mathbf{H}^T)^{-1}\mathbf{y}_j \tag{7}$$

Because of the orthogonality of full Hadamard matrix, the rows of the partial

Hadamard matrix are also orthogonal. Therefore, the partial Hadamard matrix has the property of $\mathbf{H}\mathbf{H}^T = N\mathbf{I}_M$. Substituting it to (7), the minimal $l_2$-norm least square solution to (2) is

$$\hat{\mathbf{x}}_j = \frac{\mathbf{H}^T \mathbf{y}_j}{N} \tag{8}$$

Compared with CS algorithm, (8) is easy to implement and therefore we can expect faster reconstruction speed. Rewriting $\mathbf{x}_j$ and $\mathbf{y}_j$ for ($1 \leq j \leq N$) as matrix, i.e., $\mathbf{X} = [\mathbf{x}_1, \mathbf{x}_2, \cdots \mathbf{x}_N] \in R^{N \times N}$ and $\mathbf{Y} = [\mathbf{y}_1, \mathbf{y}_2, \cdots \mathbf{y}_N] \in R^{M \times N}$, the complete STA dataset $\mathbf{X}$ can be decoded from the dataset of encoded transmissions $\mathbf{Y}$ by the following equation:

$$\mathbf{X} = \frac{\mathbf{H}^T \mathbf{Y}}{N} \tag{9}$$

In this work, we name the proposed method as minimal l2-norm least square based synthetic transmit aperture and abbreviated it as LS-STA. The reconstructed STA dataset (using LS-STA and CS-STA) and true STA dataset are beamformed using conventional delay-and-sum (DAS) method as shown in Fig. 1(d) for comparisons.

## III. Methods:

We evaluated the performance of LS-STA in simulations, phantom and in-vivo experiments by comparing it with STA and CS-STA. A simulated linear array and an L10-5 linear-array transducer (Shenzhen JiaRui Co., Shenzhen, China), with parameters listed in Table I, were used for data acquisition in the simulations and experiments, respectively.

TABLE I.  ARRAY PARAMETERS

| Parameter | Value | Unit |
|---|---|---|
| Type | L10-5 | - |
| No. of elements | 128 | - |
| Pitch | 0.3 | mm |
| Kerf | 0.03 | mm |
| Center frequency | 6.25 | MHz |
| Fractional bandwidth | 60% | - |

*A. Simulation and Experimental Setup*

We simulated the STA dataset and partial Hadamard-encoded PW dataset of a numerical phantom with Field II simulator [28]. The size of the simulated phantom is 60 × 40 × 5 mm³ (axial × lateral × elevational), with a scatterer density of 10 per resolution cell. It contains five point targets, five anechoic and five hyperechoic cylinders with radii of 3.0, 2.5, 2.0, 1.5 and 1.0 mm, located at depths of 10, 20, 30, 40 and 50 mm, as shown in Fig. 3. The average scattering amplitudes of the point targets and the hyperechoic cylinders are set to 100 and 10 times higher than that in the background. In the simulations, an acoustic attenuation coefficient of 0.5 dB/cm/MHz was set. To investigate the performance of the proposed method under noisy situations, different levels of noise (with SNRs of 0, 5 and 10 dB, respectively) were added to the simulated STA dataset. Assuming that the power of electronic noise was irrelevant to the acquisition sequence, the same noise was added to the Hadamard encoded dataset for comparison. The radio-frequency (RF) data were acquired at a sampling frequency of 100 MHz, and then down-sampled to 25 MHz.

We performed the phantom and *in-vivo* experiments using the Verasonics Vantage 256 system (Verasonics Inc., Kirkland, WA, USA) by imaging two typical views (the wire and cyst regions) of CIRS 040GSE tissue-mimicking phantom (CIRS, Norfolk, VA, USA) and the liver of a healthy volunteer with the L10-5 linear array. In the experiments, the driving voltage of transducer elements was set to 15 V. The RF data were acquired at a sampling frequency of 25 MHz.

In both the simulations and experiments, 128 STA transmissions were performed to acquire the STA dataset, and 128 Hadamard-apodized PW transmissions were performed to acquire the PW dataset. Thereafter, the received echoes corresponding to the first 16, 32, 64 and 128 PW transmissions were extracted as partial Hadamard encoded dataset for CS-STA and LS-STA reconstruction. The corresponding frame rates can be improved by 8-, 4-, 2- and 1-fold, respectively, compared with STA imaging. Their corresponding encoding matrices used in reconstruction are the first 16, 32, 64 and 128 rows of a full 128-order Hadamard matrix.

*B. Data Processing*

We reconstructed the complete STA dataset from the partial Hadamard-encoded

PW dataset with CS-STA and LS-STA in the RF domain. For CS-STA, like our previous work [22, 25], spgl1 toolbox [29, 30] was utilized to reconstruct the complete STA dataset with a tolerated error ε of 1×10⁻³ and a maximum number of iterations of 1,000.

After reconstruction, we performed DAS beamforming on the reconstructed STA dataset and true STA dataset with the same beamforming parameters (rectangular receive apodization and an f-number of 1.5) and compared their performances.

Both the reconstruction and beamforming codes were written in Matlab 2017b (The MathWorks, Inc., Natick, MA, USA) and run on Dell Precision 3450 (Intel (R) Core (TM) i7-10700 CPU @ 2.90 GHz).

## C. Evaluation Metrics

The reconstruction errors of CS-STA and LS-STA lead to degradation of the image quality. The normalized root-mean-square errors (NRMSEs) between the pre-beamformed true STA dataset and the corresponding reconstructed STA dataset using CS-STA and LS-STA (in noise-free condition) were calculated as below:

$$\text{NRMSE} = \frac{\sqrt{\frac{1}{N^e N^r N^s} \sum_{i=1}^{N^e} \sum_{j=1}^{N^r} \sum_{s=1}^{N^s} (I(e_i, r_j, s_k) - I'(e_i, r_j, s_k))^2}}{\max_{i,j,k} |I(e_i, r_j, s_k)|} \quad (10)$$

where $I$ and $I'$ are the true and reconstructed STA dataset. $N^e$, $N^r$ and $N^s$ are the total number of transmit elements, receive elements and samples, respectively. $e_i$, $r_j$, and $s_k$ are the indexes of transmit element, receive element and sample, respectively.

The lateral resolution of the proposed method was evaluated by measuring the full-width-at-half-maximum (FWHM) of the point-like targets. The generalized contrast-to-noise ratio (gCNR) [31] was calculated to evaluate the lesion detection ability of the proposed method as:

$$\text{gCNR} = 1 - \text{OVL} \quad (11)$$

where OVL denotes the overlap area between the probability density functions of the region of interest (ROI) and the background.

The SNR of the envelope images was calculated as:

$$\text{SNR} = 10 \times \log_{10}\left(\frac{\sum_{i=1}^{N^s}\sum_{j=1}^{N^e} signal(i,j)^2}{\sum_{i=1}^{N^s}\sum_{j=1}^{N^e} noise(i,j)^2}\right) \qquad (12)$$

In the simulations, the signal was obtained in the noise-free condition while the noise was the difference between the envelope images in the Gaussian noise and noise-free conditions.

## IV. Results:

A. Simulation Results

The NRMSEs between the pre-beamformed true STA dataset and the STA dataset reconstructed using CS-STA and LS-STA (in noise-free condition) with different numbers of transmissions are shown in Fig. 2. As can be observed, the NRMSEs of both CS-STA and LS-STA decrease as the number of transmissions increases, and the NRMSEs of LS-STA are the same as those of CS-STA with the same number of transmissions. Moreover, the NRMSEs of LS-STA and CS-STA are close to zero when the number of transmissions is 128, i.e., LS-STA and CS-STA reconstruct the complete STA dataset with negligible error when the number of transmissions is the same as that of STA.

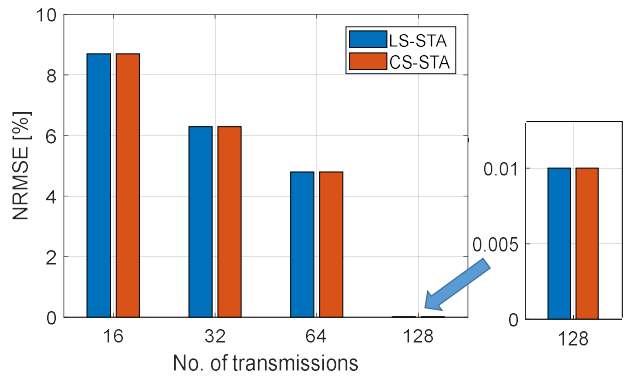

Fig. 2. NRMSEs between the true STA dataset and STA dataset reconstructed using LS-STA and CS-STA with different numbers of transmissions.

Fig. 3 shows the simulated B-mode images of different methods in the noise-free condition. It can be observed that the quality of the reconstructed hyperechoic inclusions and point targets are not sensitive to the number of transmissions for both CS-STA and LS-STA. In contrast, as the number of transmissions increases, both the CS-STA and LS-STA images present fewer artifacts inside the anechoic inclusions and become closer to the STA image. Moreover, CS-STA and LS-STA achieve similar image quality when the number of transmissions is the same.

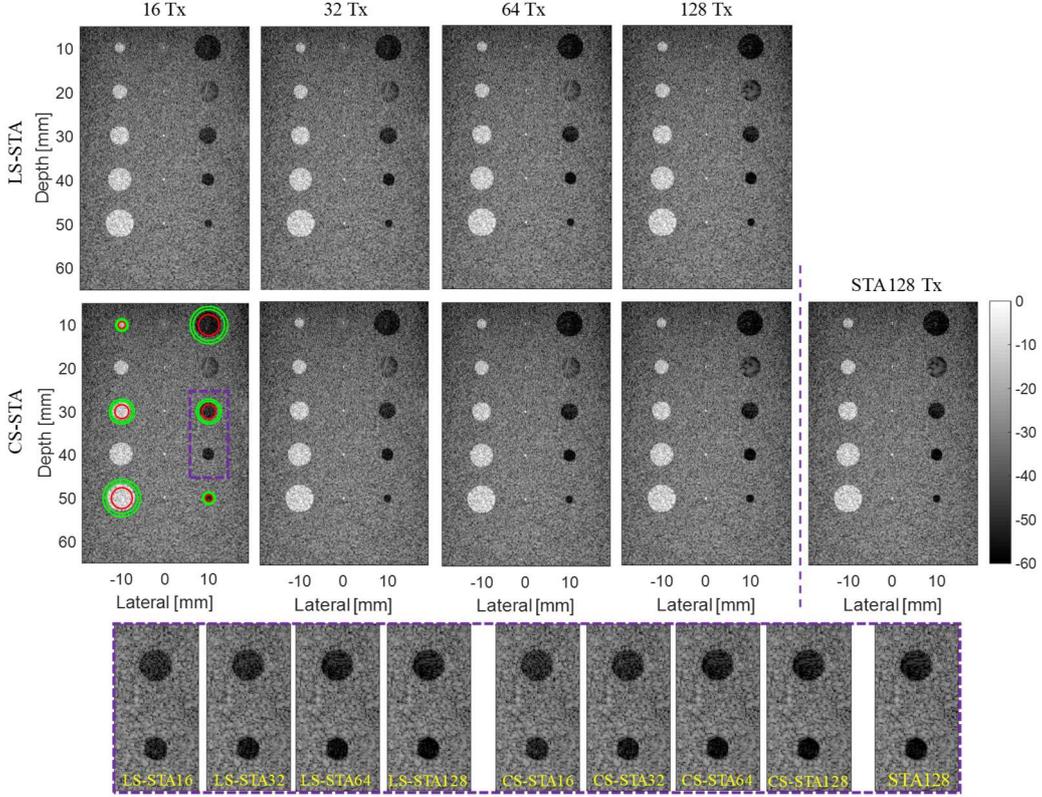

Fig. 3. B-mode images reconstructed using LS-STA (the first row) and CS-STA (the second row) with 16, 32, 64 and 128 transmissions (from the first to the fourth columns) and STA with 128 transmissions in noise-free condition. The red circles and green concentric circles indicate the ROIs and the corresponding backgrounds for the calculations of gCNRs. The figures in the bottom row are the zoomed-in versions of the selected regions indicated by the purple dotted box.

Fig. 4 presents the simulated B-mode images of different methods at SNR of 0 dB. The STA image severely suffers from the noise, and presents high-level artifacts inside the anechoic inclusions. In contrast, thanks to the high-energy Hadamard-encoded PW transmissions, the reconstructed STA dataset has a higher SNR, compared with the true STA dataset. Therefore, CS-STA and LS-STA with 16 transmissions achieve better visual quality than STA with 128 transmissions in terms of contrast between the anechoic cysts and background. Moreover, as the number of PW transmissions increases, the artifacts inside the anechoic cysts can be suppressed to further improve the contrast, as shown in the zoomed-in images.

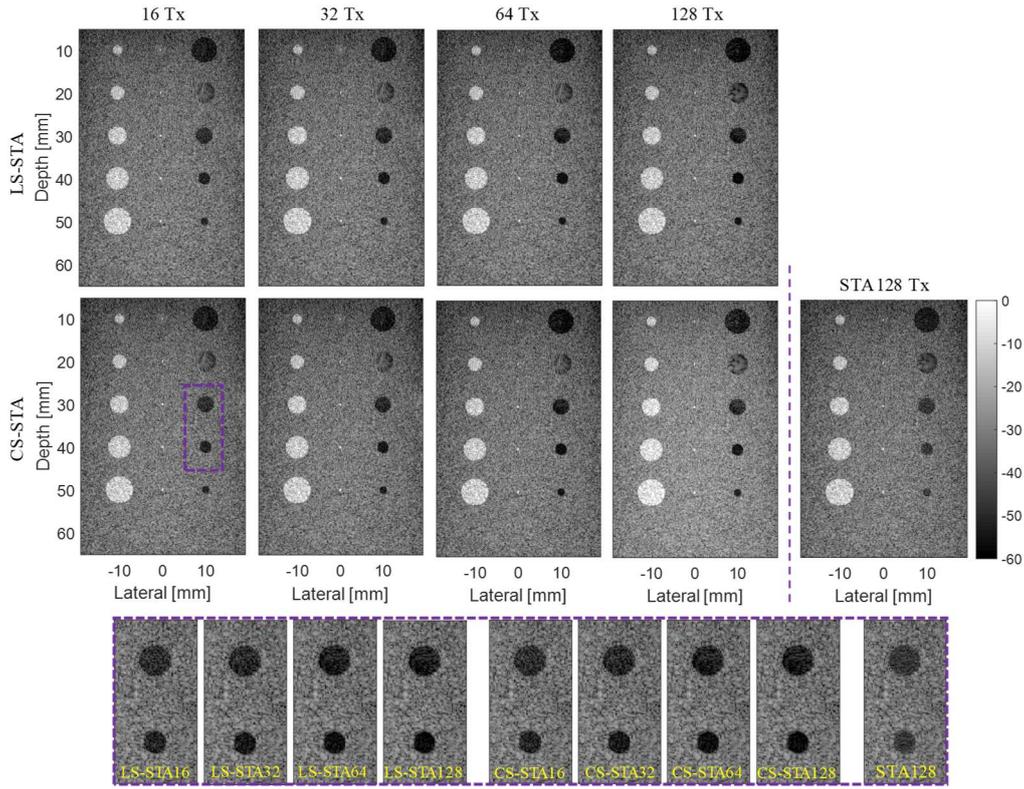

Fig. 4. B-mode images reconstructed using LS-STA (the first row) and CS-STA (the second row) with 16, 32, 64 and 128 transmissions (from the first to the fourth columns) and STA with 128 transmissions at SNR of 0 dB. The figures in the bottom row are the zoomed-in versions of the selected regions indicated by the purple dotted box.

Fig. 5 presents the quantitative assessments of CS-STA and LS-STA images with different numbers of transmissions (16, 32, 64 and 128) at different noise levels (0 dB, 5 dB, 10 dB and noise-free). The FWHMs of all the five point targets are measured to evaluate the lateral resolutions of different methods. It can be observed that CS-STA and LS-STA achieve almost the same results with the same number of transmissions, in terms of both gCNRs and FWHMs. For the calculation of gCNR, as demonstrated in Fig. 3, the red circles (with radii being 0.8 times of the radii of the corresponding cylinders) indicate the ROIs, while the regions between the green concentric circles (with inner and outer radii being 1.2 and 1.44 times of the radii of the corresponding cylinders, respectively) are the background. The gCNRs of the hyperechoic targets, axial and lateral FWHMs of the point targets of images reconstructed using CS-STA and LS-STA are not sensitive to the number of transmissions, and are very close to those obtained using STA with 128 transmissions, as shown in Fig. 5(b)-(d). The FWHM results verify that LS-STA has the capability of maintaining the spatial resolution of STA, as CS-STA does. In contrast, the comparisons on the gCNRs of the anechoic targets are heavily affected by the noise level. As shown in Fig. 5(a), CS-STA and LS-STA are robust to noise and achieve close results at different noise levels. In contrast, STA is very sensitive to noise and the obtained gCNR decreases with the SNR. Therefore, it can be observed that the gCNRs of STA are comparable/worse than those of CS-STA and LS-STA (with different numbers of transmissions), at SNRs of 0, 5,10 dB and noise-free condition, respectively. These results demonstrate the superiority of the proposed LS-STA method in low-SNR conditions, compared with STA.

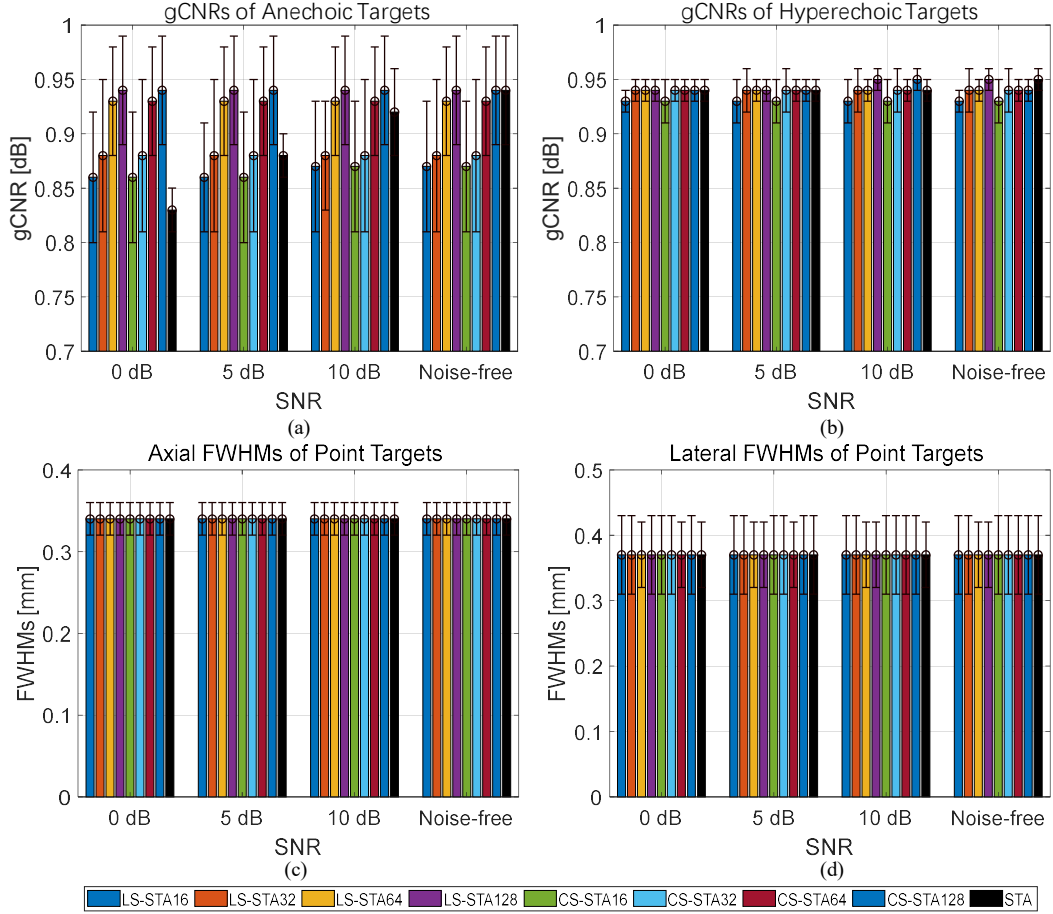

Fig. 5. The means and deviations of (a) gCNRs of the marked anechoic targets, (b) gCNRs of the marked hyperechoic targets, (c) axial FWHMs of five point targets and (d) lateral FWHMs of five point targets, with respect to LS-STA, CS-STA and STA at different SNRs.

Table II presents the reconstruction time of CS-STA and LS-STA with different numbers of transmissions. As shown, the average reconstruction time of CS-STA is around 1,450 s for different numbers of transmissions. As a comparison, the average reconstruction time of LS-STA is around 0.30 s for different numbers of transmissions. Although this reconstruction time is not optimal because all the codes are written and run in Matlab, there is no doubt that with minimal $l_2$-norm least square method, LS-STA is capable of accelerating the reconstruction significantly, compared with CS-STA.

TABLE II. RECONSTRUCTION TIME OF CS-STA AND LS-STA WITH DIFFERENT NUMBERS OF TRANSMISSIONS

| No. of transmissions | CS-STA [s] | LS-STA [s] | Acceleration |
|---|---|---|---|
| 16 | 1382 | 0.27 | 5,119× |
| 32 | 1400 | 0.27 | 5,185× |
| 64 | 1435 | 0.31 | 4,629× |
| 128 | 1573 | 0.37 | 4,251× |

The average SNRs of reconstructed envelope images with different methods were calculated for comparison. CS-STA and LS-STA achieve the same SNRs when the number of transmissions is the same. STA with 128 transmissions achieves SNR of 17.3 dB, while CS-STA and LS-STA with 16, 32, 64 and 128 transmissions achieve SNRs of 18.1, 21.0, 25.1 and 38.3 dB, respectively. In summary, thanks to the high-energy Hadamard-encoded transmissions, the reconstructed STA dataset from a small number of PW transmissions has higher SNR than the true STA dataset, resulting in an improved SNR for beamformed images. In addition, the SNR improvement of CS-STA and LS-STA over STA increases with the number of transmissions. As analyzed in [32], the SNR improvement brought by Hadamard encoded transmission is $\sqrt{M}$, in which $M$ is the number of encoded transmissions.

B. Experiments on Phantoms

Fig. 6 presents the B-mode images for the wire region of the tissue-mimicking phantom reconstructed using different methods. As shown, LS-STA and CS-STA obtain very similar image quality to that of STA. The lateral FWHMs of two wire targets (indicated by the red arrows) are calculated to compare the spatial resolution of different methods and shown in Fig. 6. These results are in accordance with the simulation results, and further verify the capability of LS-STA in maintaining the resolution of STA. In addition, the calculated gCNRs of the hyperechoic inclusions are very close for different methods. To investigate the repeatability and robustness of the proposed method, the gCNR and FWHMs were evaluated on five consecutive frames with the same ROI and wire targets, respectively. The standard deviations of the gCNRs of all the methods are less than 0.0025. The standard deviations of the FWHMs of all the methods are less than 0.005. These results verify the robustness of the proposed method against acquisition noise.

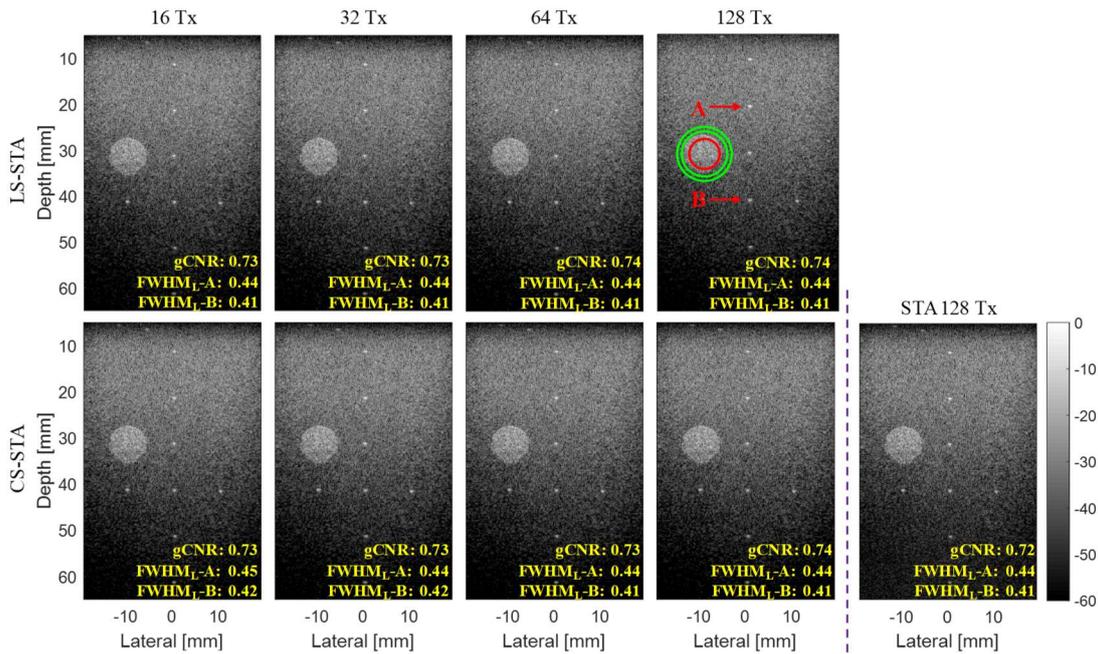

Fig. 6. B-mode images for the wire region of the phantom reconstructed using LS-STA (the first row) and CS-STA (the second row) with 16, 32, 64 and 128 transmissions (from the first to the fourth columns) and STA with 128 transmissions. The red arrows indicate the wire targets used for the measurements of FWHMs. The red circle and the concentric green circles indicate the ROI and background used for the calculations of gCNR.

Fig. 7 presents the B-mode images of the cyst region of the phantom reconstructed using different methods. As shown, severe noise can be observed in the anechoic cysts in the image obtained using STA. In contrast, the noise is effectively suppressed in the images obtained using LS-STA and CS-STA methods, especially for ROI C. To quantify the image contrast, the gCNRs of three ROIs and the corresponding backgrounds (with the same area), which were indicated by the red and green circles, respectively, are calculated and shown in Fig. 8. It can be observed that STA achieves the highest and the worst gCNRs for ROIs A and C, respectively, compared with LS-STA and CS-STA. The contrasts of the anechoic inclusions (ROIs A and C) can be improved by increasing the number of transmissions, while the contrast of the hypoechoic inclusion (ROI B) and the FWHM of the wire target are not very sensitive to the number of transmissions. To investigate the repeatability and robustness of the proposed method, the gCNRs and FWHMs were evaluated on five consecutive frames with the same ROI and wire targets, respectively. The standard deviations of the gCNRs of all the methods for the three ROIs are less than 0.003. The standard deviations of the FWHMs of all the methods for the wire target are less than 0.003. These results verify the robustness of the proposed method against acquisition noise.

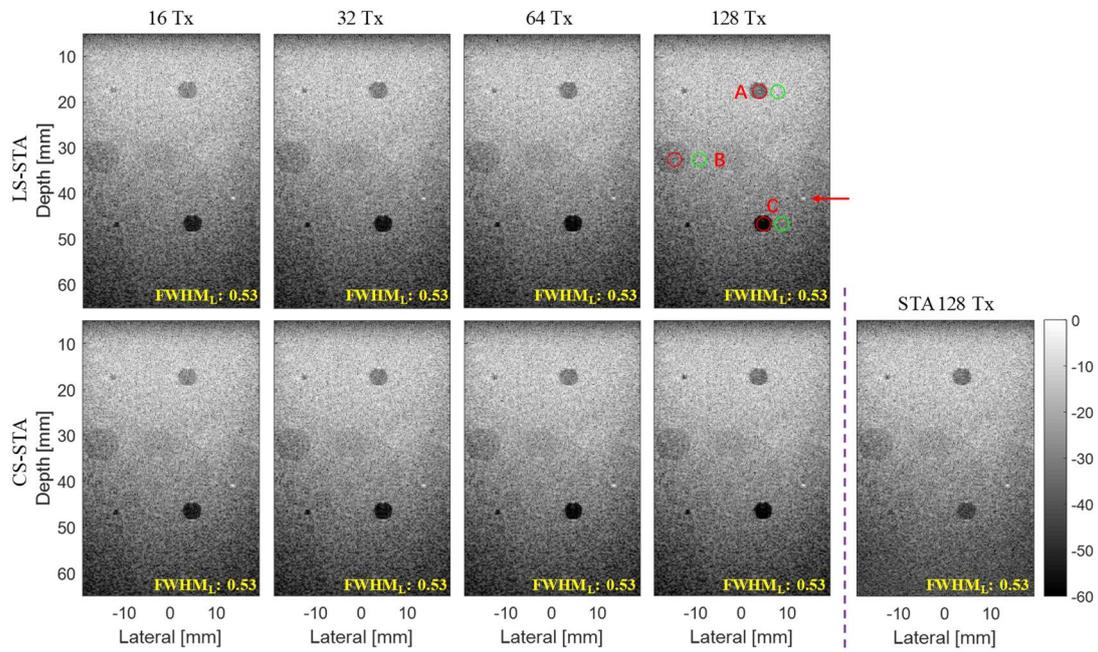

Fig. 7. B-mode images for the cyst region of the phantom reconstructed using LS-STA (the first row) and CS-STA (the second row) with 16, 32, 64 and 128 transmissions (from the first to the fourth columns) and STA with 128 transmissions. The red and green circles indicate the ROIs and backgrounds used for the calculations of gCNRs. The red arrow indicates the wire target used for the measurements of FWHM.

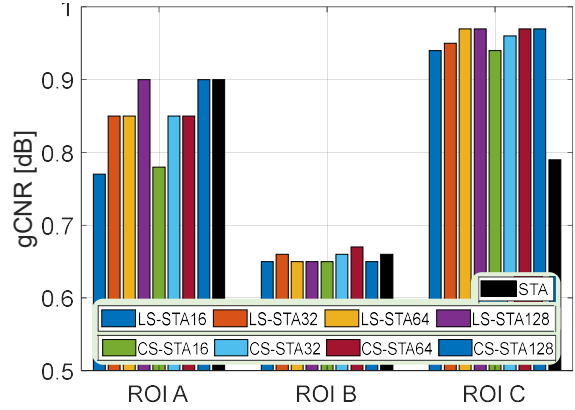

Fig. 8. Calculated gCNRs of the three ROIs.

C. In-vivo Results

Fig. 9 shows the B-mode images for the liver of a healthy volunteer reconstructed using different methods. Severe noise can be observed in the STA image, resulting in a very low contrast for vessel lumens. As indicated by the red arrows in the STA image, the vessel is not discriminable from the surrounding high-level noise. Using Hadamard-encoded PW transmissions, LS-STA and CS-STA achieve higher image quality, especially in terms of contrast, and the vessel lumens are clearly visible. A tissue fiber is selected (indicated by the red line) to measure the axial FWHMs as spatial resolutions of different methods. As shown in Fig. 9, LS-STA and CS-STA with different numbers of transmissions achieve the same FWHMs as STA. To quantify the contrast between the vessel lumen and the background, ROI (indicated by the red circle) and background (indicated by the green circle) with radii of 1.5 mm are selected to calculate the gCNRs of different methods. To investigate the repeatability and robustness of the proposed method, the quantitative evaluations of gCNRs were performed on five consecutive frames with a fixed ROI. The means and standard deviations of the measured gCNRs for different methods are shown in Table III. Note that the relatively large standard deviations may because of the tissue motion. Results demonstrate that CS-STA and LS-STA with only 16 transmissions can achieve significantly improved gCNRs, compared with STA. In addition, the gCNRs of LS-STA and CS-STA increase as the number of transmissions increases from 16 to 128.

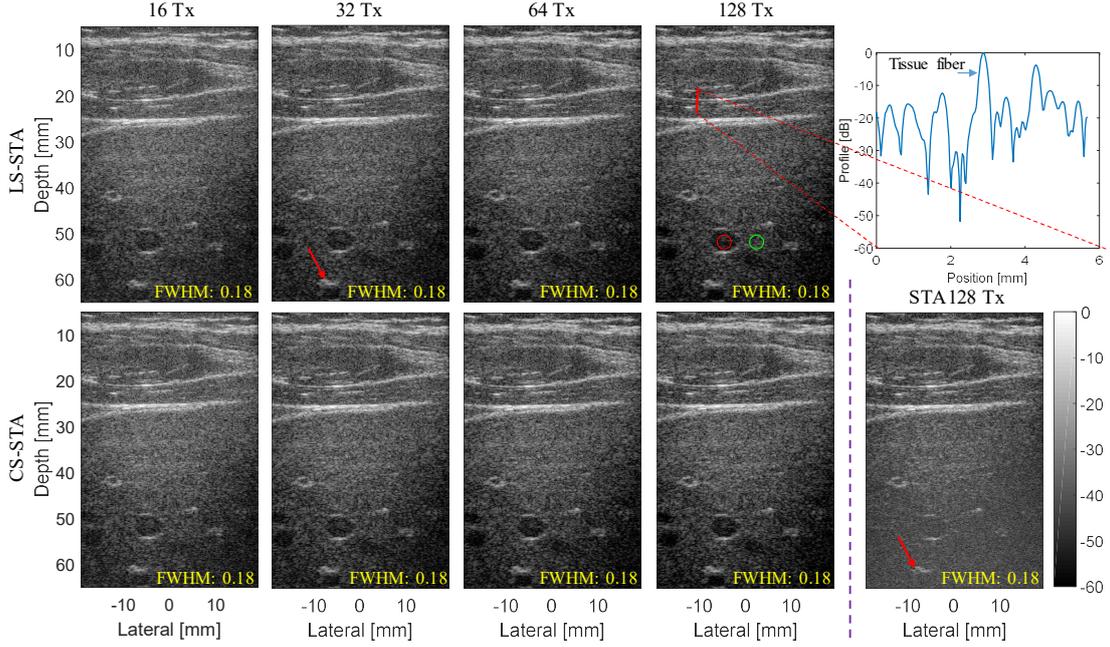

Fig. 9. B-mode images for the liver of a healthy human subject reconstructed using LS-STA (the first row) and CS-STA (the second row) with 16, 32, 64 and 128 transmissions (from the first to the fourth columns) and STA with 128 transmissions. The red arrows indicate a vessel lumen which is not discriminable from the background noise in the STA image, but is clearly visible in the LSSTA and CSSTA images. The red line indicates the selected tissue fiber for the measurements of axial FWHMs, and its cross-sectional profile is plotted on the top-right corner.

TABLE III. MEANS AND STANDARD DEVIATIONS OF MEASURED gCNRs FOR SELECTED ROI IN LIVER IMAGES

| Method | 16 Tx | 32 Tx | 64 Tx | 128 Tx |
| --- | --- | --- | --- | --- |
| LS-STA | 0.34 ± 0.04 | 0.41 ± 0.03 | 0.49 ± 0.04 | 0.51 ± 0.05 |
| CS-STA | 0.34 ± 0.04 | 0.40 ± 0.03 | 0.49 ± 0.04 | 0.51 ± 0.05 |
| STA | -- | -- | -- | 0.19 ± 0.01 |

## V. Discussions:

### A. Explanation on the results

The simulation, phantom and *in-vivo* results demonstrate that with the minimal $l_2$-norm least square method, LS-STA achieves the same reconstruction performance as CS-STA in terms of reconstruction error (Fig. 2) and image quality (Figs. 3-9). In addition, the NRMSE between the STA datasets reconstructed with LS-STA and CS-STA in the simulations (at SNR of 0 dB) are calculated as $3.3 \times 10^{-5}$. As shown in Figs. 5(c), 5(d), 6, 7 and 9, the proposed LS-STA can maintain the high resolution of STA (i.e., the most important property of STA) with a small number of transmissions (low to 16 Tx), as CS-STA does. In addition, when in low-SNR condition (Figs. 4 and 5) or in deep region (Figs. 7 and 9), CS-STA and LS-STA can achieve much higher contrast (gCNR) than STA, with a small number of transmissions (low to 16 Tx). Moreover, LS-STA (~0.3s) achieves over ~5×10³ times faster reconstruction speed than CS-STA (~1,450 s) which utilizes CS algorithm. It should be noted that the proposed LS-STA was currently implemented in Matlab and run on a central processing unit (CPU). Parallel computation using a graphics processing unit (GPU) can further accelerate the reconstruction and pave the way for its real-time application.

The quality of image reconstructed using LS-STA is determined comprehensively by the reconstruction error and SNR of the reconstructed STA dataset. Compared with STA, LS-STA improves the SNR of the reconstructed STA dataset by transmitting high-energy Hadamard-encoded PW, but also introduces reconstruction errors into the reconstructed STA dataset (Fig. 2). Therefore, for the anechoic cyst (ROI A) (Fig. 7) located in shallow regions, from which the STA sequence can acquire dataset with an acceptable SNR, the image quality of LS-STA (with fewer transmissions) is typically worse than that of STA. When the number of LS-STA/CS-STA transmissions is equal to the number of STA transmissions, the reconstruction error decreases to nearly zero, as demonstrated by Fig. 2, and LS-STA and CS-STA achieve better image quality than STA. In contrast, LS-STA can achieve higher contrast than STA for the anechoic cyst (ROI C, Fig. 7) and liver (Fig. 9) located in deep regions. These results suggest the superiority of LS-STA in low-SNR condition.

For least-square solution $\hat{x} = H^T(HH^T)^{-1}y$, the variance of $\hat{x}$ can be obtained as a function of $\sigma_y^2$ as $var(\hat{x}_j) = \sigma_y^2[(HH^T)^{-1}]_{jj}$, where the $[(HH^T)^{-1}]_{jj}$ denotes the j$^{th}$ diagonal element of $(HH^T)^{-1}$ [33]. Therefore, considering $HH^T = NI_M$, the variation of the solution estimated using the proposed method is only 1/$N$ of the variation of acquisition noise. That is, the proposed method is relatively robust to noise.

The results demonstrate that the image quality of LS-STA and CS-STA improves as the number of transmissions increases. However, it should be noted that more transmissions would in turn lead to a decreased frame rate. Tissue motion is another factor which may affect the reconstruction accuracy, as the linear combination relationship is held in static acquisition situation. In summary, in real applications, the trade-off between these above factors should be taken into consideration to determine the number of transmissions for LS-STA.

It can be observed from Figs. 7 and 8 that the contrasts of ROI A are lower than those of ROI C. This degradation can be attributed to the grating lobe artifacts generated by the used λ-pitch array (λ = 0.25 mm, pitch = 0.3 mm) [34].

In our recent study, CS-STA was extended from RF domain to in-phase/quadrature (IQ) domain to accelerate the STA dataset reconstruction, by reducing the amount of data to be processed [35]. More specifically, the I and Q components of the STA dataset are reconstructed using CS algorithm individually. LS-STA can also be extended to IQ domain for acceleration, but without the need to reconstruct two quadrature components separately, resulting in a halved computational cost. The IQ-domain CS-STA and LS-STA were used to reconstruct the B-mode images (Fig. 10) of the simulated phantom at an SNR of 0 dB for validation. More specifically, the simulated RF dataset was demodulated and down-sampled at a factor of 6 to obtain the IQ data for reconstruction. The quantitative results (Fig. 11) demonstrate that IQ-domain CS-STA and LS-STA achieve very close results with the same number of transmissions.

The main difference between CS-STA and LS-STA is that they apply different objective functions on the signal to be reconstructed $\mathbf{x}_j$. CS-STA assumes that $\mathbf{x}_j$ can be sparsely represented with $\mathbf{v}_j$, i.e., most entries of $\mathbf{v}_j$ are zero (or close to zero). To obtain $\mathbf{v}_j$ numerically, CS-STA finds $\mathbf{v}_j$ which has the minimal $l_1$-norm subject to $\|\mathbf{y}_j - \mathbf{H}\mathbf{\Psi}\mathbf{v}_j\|_2 \le \varepsilon$. This problem has no analytical solutions and requires a number of iteration steps. The iteration process is time-consuming. In comparison, LS-STA finds $\mathbf{x}_j$ which has the minimal $l_2$-norm subject to $\mathbf{y}_j = \mathbf{H}\mathbf{x}_j$. This problem has an analytical solution. Besides, because the Hadamard matrix is orthogonal, the solution becomes $\hat{\mathbf{x}}_j = \mathbf{H}^T \mathbf{y}_j / N$ and it only needs to perform a matrix-multiplication to obtain the analytical solution. Therefore, the reconstruction process of LS-STA can be easily implemented and the corresponding process time is significantly shorter than that of CS-STA.

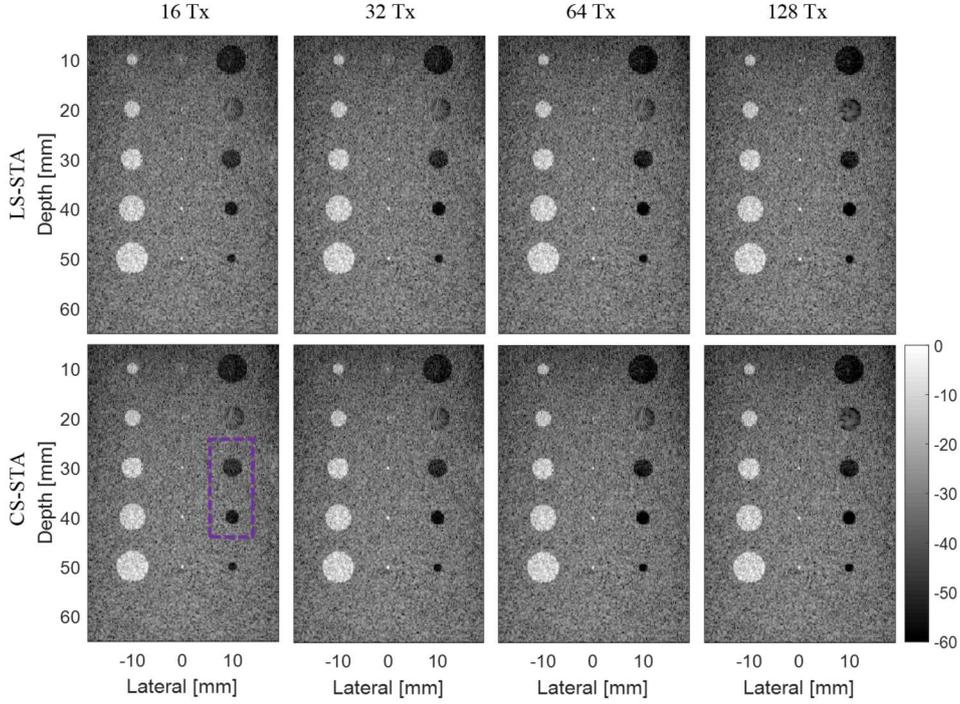

Fig. 10. B-mode images reconstructed in IQ domain using LS-STA (the first row) and CS-STA (the second row) with 16, 32, 64 and 128 transmissions (from the first to the fourth columns). The ROIs and wire targets selected for quantitative evaluation are the same as those indicated in Fig. 3.

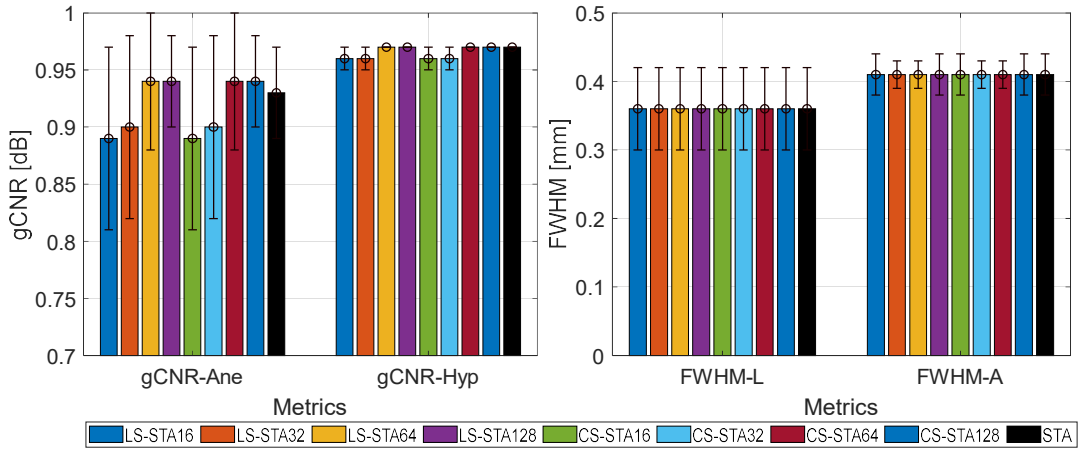

Fig. 11. The means and standard deviations of gCNRs of the anechoic and hyperechoic targets, and axial/lateral FWHMs for the wire targets (reconstructed in IQ domain).

## B. Limitation and Future work

CS-STA and LS-STA apply different objective functions on the signal $\mathbf{x}_j$ to be reconstructed. As discussed before, LS-STA can obtain the same reconstruction results as CS-STA. This may be explained by that the reconstructed $\mathbf{x}_j$ satisfies these two constraints at the same time because these two constraints are not contradictory. However, we should be aware that the minimization function of the proposed LS-STA method lacks an error tolerance. In consequence, noise may be fitted in the solution, resulting in a noisy reconstructed STA dataset and the corresponding degraded image quality. For this problem, Tikhonov regularization [36] may be a very useful tool to achieve a balance between the error tolerance and the data fitting. Nevertheless, there is one feature about the complete STA dataset $\mathbf{X}$. As Section II presents, each entry $x_{n,j}$ of $\mathbf{X}$ denotes the received backscattered echo received by the $j^{\text{th}}$ element when the $n^{\text{th}}$ element is activated. According to the acoustic reciprocity, $x_{n,j}$ equals $x_{j,n}$ in the noise-free condition. If the noise level is low, $\mathrm{x}_{n,j}$ would be quite close to $\mathrm{x}_{j,n}$. Therefore, the complete STA dataset $\mathbf{X}$ is approximately a symmetrical matrix if the SNR is relatively high. Fig. 12 shows an example of the STA data in the transmit-receive space. We can find that the simulated complete STA dataset in the noise-free condition is symmetrical. For the noisy condition, the complete STA dataset is still approximately symmetrical. In this work, we did not utilize this feature. We may make use of it to improve the reconstruction accuracy in the future.

In this work, we implemented the Hadamard encoded transmission by applying each row of the partial Hadamard matrix as the transmit apodization and all the elements are activated at the same time. In [10], Bottenus applied transmit delays to transmit focused wave and formulated the linear relationship between the complete STA dataset and focused wave dataset in the frequency domain. The coding matrix $\mathbf{H}$ is therefore a complex matrix and is determined by the applied time delays. Bottenus assumed $\mathbf{HH}^H$ is an identity matrix and reconstructed the complete STA dataset from the focused wave dataset in the frequency domain with an equation similar to (8). Because this assumption is not satisfied well when transmitting fewer focused waves, this method could not improve the frame rate significantly. This may be combined with LS-STA, i.e., activating all the elements with different apodization and delays to make $\mathbf{HH}^H$ close to an identity matrix as much as possible.

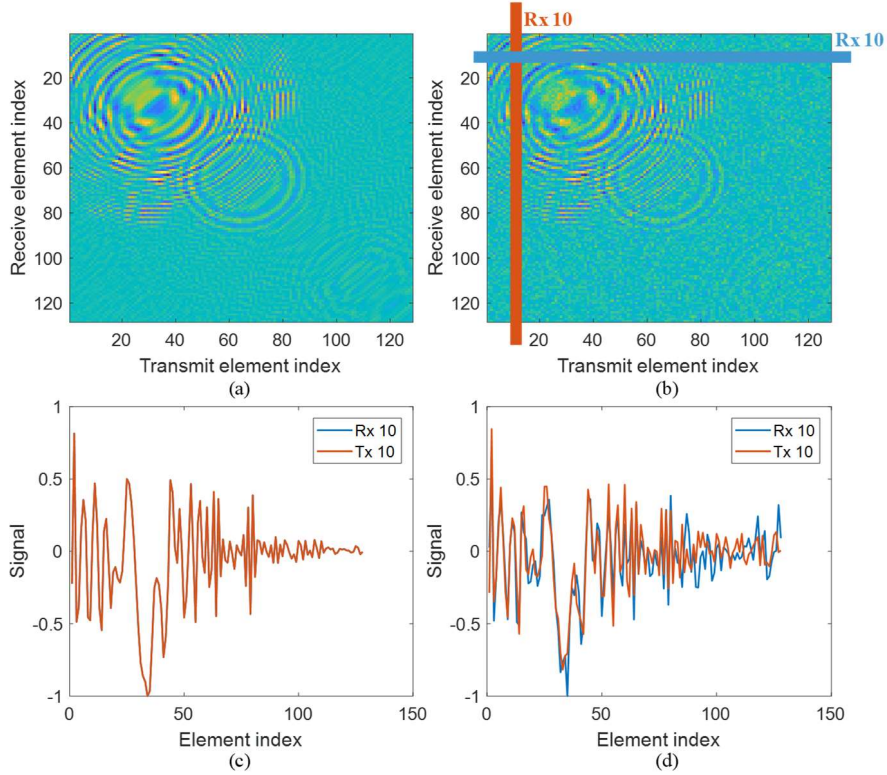

Fig. 12. True full STA dataset from (a) simulations in the noise-free condition and (b) simulations at SNR of 0 dB. The echoes received by the 10th element when all the elements are activated sequentially (Rx 10) and echoes received by all the elements when the 10th element is activated (Tx 10) for noise-free and noisy (SNR of 0 dB) situations are plotted in (c) and (d), respectively. We can find that the full STA dataset is symmetrical in noise-free condition.

The key reason for the fast reconstruction of the proposed method is that the encoding matrix is orthogonal, which makes the inverse of $\mathbf{HH}^T$ be a scaled identity matrix. As mentioned above, other encoding matrices which are not orthogonal may be applied in some cases. For example, the optimized transmit apodization obtained using deep learning method named ApodNet [37]. In these cases, the proposed method needs to calculate the inverse of $\mathbf{HH}^T$. Nevertheless, we need to calculate the inverse of $\mathbf{HH}^T$ only once because the encoding matrix is the same for all samples. Therefore, the reconstruction time would not increase significantly. In addition, it may be pre-calculated and stored in the RAM. For some encoding matrices $\mathbf{H}$, the matrix $\mathbf{HH}^T$ is not full-rank. Therefore, we need to calculate the inverse of $(\mathbf{HH}^T + \varepsilon \mathbf{I})$ in which $\varepsilon$ is Tikhonov regularization parameter. Such regularization has been applied in ultrasound imaging to reconstruct beamformed image from channel data [36].

It is well known that the coherent compounding procedure in the STA

beamforming process would produce motion artifacts. In this work, we find that the image quality would improve when the number of transmissions increases if the object is static (simulations and phantom experiments) or the motion is not fast (*in-vivo* experiments). However, the image quality would deteriorate if the object moves fast. Therefore, it is necessary to assess the image quality of LS-STA with different numbers of transmissions for a moving object, which is our ongoing work.

## VI. Conclusion:

In this work, we utilize the minimal $l_2$-norm least square method to reconstruct the complete STA dataset from the dataset produced by partial Hadamard-encoded PW transmissions. Thanks to the orthogonality of Hadamard matrix, this method is easy to implement. Compared with CS-STA, the proposed method is capable of accelerating the reconstruct speed over $\sim 5\times 10^3$ times with the same reconstruction accuracy. Compared with STA, the proposed method is capable of achieving higher contrast and SNR with maintained spatial resolution. These advantages pave its way to real applications.

## Acknowledgements:

This work was supported by National Natural Science Foundation Grants of China (61871251, 11874382), Shenzhen Research Grant (GJHZ20180420180920529, JCYJ20170817171836611, and ZDSYS201802061806314), Shenzhen Double Chain Project [2018]256, Youth Innovation Promotion Association CAS, CAS research projects (QYZDB-SSW-JSC018) and Guangdong Special Support Program.